\documentclass[prb,aps,amsmath,twocolumn,eqsecnum]{revtex4}

    \usepackage{graphicx,amssymb,bm}

\usepackage{epic}

\newcommand\beq{\begin{equation}}
\newcommand\eeq{\end{equation}}
\newcommand\beqa{\begin{eqnarray}}
\newcommand\eeqa{\end{eqnarray}}
\newcommand{\dd}{\text{d}}
\newcommand{\nn}{\nonumber\\}

\newcommand{\rr}{\mathbf{r}}

\newcommand{\py}{\text{PY}}
\newcommand{\hnc}{\text{HNC}}
\newcommand{\ex}{\text{exact}}
\newcommand{\app}{\text{app}}

\newcommand{\vi}{\text{v}}
\newcommand{\co}{\text{c}}
\newcommand{\en}{\text{e}}

\newcommand{\xx}{5}
\newcommand{\yy}{10}
\newcommand{\yyh}{5}

\newcommand{\xmax}{40}
\newcommand{\ymax}{22}

\newcommand{\yone}{\begin{picture}(\xmax,\ymax)(-\xx,\yy)
\setlength{\unitlength}{.1mm} \put(50,100){\circle*{10}}
\put(100,0){\circle{10}} \put(0,0){\circle{10}}
\put(50,100){\line(1,-2){48}} \put(50,100){\line(-1,-2){48}}
\end{picture}}

\newcommand{\ytwoA}{\begin{picture}(\xmax,\ymax)(-\xx,\yy)
\setlength{\unitlength}{.1mm} \put(0,100){\circle*{10}}
\put(100,100){\circle*{10}} \put(100,0){\circle{10}}
\put(0,0){\circle{10}} \put(0,100){\line(100,0){100}}
\put(0,100){\line(0,-100){95}} \put(100,100){\line(0,-100){95}}
\end{picture}}

\newcommand{\ytwoB}{\begin{picture}(\xmax,\ymax)(-\xx,\yy)
\setlength{\unitlength}{.1mm} \put(0,0){\circle{10}}\put(0,100){\circle*{10}}
\put(100,100){\circle*{10}} \put(100,0){\circle{10}}
 \put(0,100){\line(100,0){100}}
\put(0,100){\line(0,-100){95}} \put(100,100){\line(0,-100){95}}
\put(100,100){\line(-1,-1){95}}
\end{picture}}

\newcommand{\ytwoC}{\begin{picture}(\xmax,\ymax)(-\xx,\yy)
\setlength{\unitlength}{.1mm} \put(0,100){\circle*{10}}
\put(100,100){\circle{10}} \put(100,0){\circle*{10}}
\put(0,0){\circle{10}} \put(0,100){\line(100,0){95}}
\put(0,100){\line(0,-100){95}} \put(100,0){\line(0,100){95}}
\put(100,0){\line(-100,0){95}}
\end{picture}}

\newcommand{\ytwoCbis}{\begin{picture}(\xmax,\ymax)(-\xx,\yy)
\setlength{\unitlength}{.1mm} \put(0,100){\circle*{10}}
\put(100,100){\circle*{10}} \put(100,0){\circle{10}}
\put(0,0){\circle{10}} \put(0,100){\line(0,-100){95}}
\put(100,100){\line(0,-100){95}} \put(100,100){\line(-1,-1){95}}
\put(0,100){\line(1,-1){95}}
\end{picture}}

\newcommand{\ytwoD}{\begin{picture}(\xmax,\ymax)(-\xx,\yy)
\setlength{\unitlength}{.1mm} \put(0,100){\circle*{10}}
\put(100,100){\circle*{10}} \put(100,0){\circle{10}}
\put(0,0){\circle{10}} \put(0,100){\line(100,0){100}}
\put(0,100){\line(0,-100){95}} \put(100,100){\line(0,-100){95}}
\put(100,100){\line(-1,-1){95}} \put(0,100){\line(1,-1){95}}
\end{picture}}

\newcommand{\chitwo}{\begin{picture}(\xmax,\ymax)(-\xx,\yyh)
\setlength{\unitlength}{.1mm}
\put(100,25){\circle*{10}} \put(0,25){\circle{10}}
\put(5,25){\line(1,0){100}}
\end{picture}}

\newcommand{\yoneA}{\begin{picture}(\xmax,\ymax)(-\xx,\yy)
\setlength{\unitlength}{.1mm} \put(50,100){\circle*{10}}
\put(100,0){\circle*{10}} \put(0,0){\circle{10}}
\put(50,100){\line(1,-2){48}} \put(50,100){\line(-1,-2){47}}
\end{picture}}

\newcommand{\yones}{\begin{picture}(\xmax,\ymax)(-\xx,\yy)
\setlength{\unitlength}{.1mm} \put(50,100){\circle*{10}}
\put(100,0){\circle*{10}} \put(0,0){\circle{10}}
\put(50,100){\line(1,-2){48}} \put(50,100){\line(-1,-2){47}}
\put(5,0){\line(1,0){100}}
\end{picture}}

\newcommand{\ytwoAA}{\begin{picture}(\xmax,\ymax)(-\xx,\yy)
\setlength{\unitlength}{.1mm} \put(0,100){\circle*{10}}
\put(100,100){\circle*{10}} \put(100,0){\circle*{10}}
\put(0,0){\circle{10}} \put(0,100){\line(100,0){100}}
\put(0,100){\line(0,-100){95}} \put(100,100){\line(0,-100){95}}
\end{picture}}

\newcommand{\ytwoAd}{\begin{picture}(\xmax,\ymax)(-\xx,\yy)
\setlength{\unitlength}{.1mm} \put(0,100){\circle*{10}}
\put(100,100){\circle*{10}} \put(100,0){\circle*{10}}
\put(0,0){\circle{10}} \put(0,100){\line(100,0){100}}
\put(0,100){\line(0,-100){95}} \put(100,100){\line(0,-100){95}}
\dashline{20}(5,0)(100,0)
\end{picture}}

\newcommand{\ytwoAs}{\begin{picture}(\xmax,\ymax)(-\xx,\yy)
\setlength{\unitlength}{.1mm} \put(0,100){\circle*{10}}
\put(100,100){\circle*{10}} \put(100,0){\circle*{10}}
\put(0,0){\circle{10}} \put(0,100){\line(100,0){100}}
\put(0,100){\line(0,-100){95}} \put(100,100){\line(0,-100){95}}
\put(5,0){\line(100,0){100}}
\end{picture}}

\newcommand{\ytwoBB}{\begin{picture}(\xmax,\ymax)(-\xx,\yy)
\setlength{\unitlength}{.1mm} \put(0,0){\circle{10}}\put(0,100){\circle*{10}}
\put(100,100){\circle*{10}} \put(100,0){\circle*{10}}
 \put(0,100){\line(100,0){100}}
\put(0,100){\line(0,-100){95}} \put(100,100){\line(0,-100){95}}
\put(100,100){\line(-1,-1){95}}
\end{picture}}

\newcommand{\ytwoBd}{\begin{picture}(\xmax,\ymax)(-\xx,\yy)
\setlength{\unitlength}{.1mm} \put(0,0){\circle{10}}\put(0,100){\circle*{10}}
\put(100,100){\circle*{10}} \put(100,0){\circle*{10}}
 \put(0,100){\line(100,0){100}}
\put(0,100){\line(0,-100){95}} \put(100,100){\line(0,-100){95}}
\put(100,100){\line(-1,-1){95}}
\dashline{20}(5,0)(100,0)
\end{picture}}

\newcommand{\ytwoBdd}{\begin{picture}(\xmax,\ymax)(-\xx,\yy)
\setlength{\unitlength}{.1mm} \put(0,0){\circle*{10}}\put(0,100){\circle{10}}
\put(100,100){\circle*{10}} \put(100,0){\circle*{10}}
 \put(5,100){\line(100,0){100}}
\put(0,95){\line(0,-100){95}} \put(100,100){\line(0,-100){95}}
\put(100,100){\line(-1,-1){100}}
\dashline{20}(5,0)(100,0)
\end{picture}}

\newcommand{\ytwoBs}{\begin{picture}(\xmax,\ymax)(-\xx,\yy)
\setlength{\unitlength}{.1mm} \put(0,0){\circle{10}}\put(0,100){\circle*{10}}
\put(100,100){\circle*{10}} \put(100,0){\circle*{10}}
 \put(0,100){\line(100,0){100}}
\put(0,100){\line(0,-100){95}} \put(100,100){\line(0,-100){95}}
\put(100,100){\line(-1,-1){95}}
\put(5,0){\line(100,0){100}}
\end{picture}}

\newcommand{\ytwoBss}{\begin{picture}(\xmax,\ymax)(-\xx,\yy)
\setlength{\unitlength}{.1mm} \put(0,0){\circle*{10}}\put(0,100){\circle{10}}
\put(100,100){\circle*{10}} \put(100,0){\circle*{10}}
 \put(5,100){\line(100,0){100}}
\put(0,95){\line(0,-100){95}} \put(100,100){\line(0,-100){95}}
\put(100,100){\line(-1,-1){95}}
\put(5,0){\line(100,0){100}}
\end{picture}}

\newcommand{\ytwoDd}{\begin{picture}(\xmax,\ymax)(-\xx,\yy)
\setlength{\unitlength}{.1mm} \put(0,100){\circle*{10}}
\put(100,100){\circle*{10}} \put(100,0){\circle*{10}}
\put(0,0){\circle{10}} \put(0,100){\line(100,0){100}}
\put(0,100){\line(0,-100){95}} \put(100,100){\line(0,-100){95}}
\put(100,100){\line(-1,-1){95}} \put(0,100){\line(1,-1){95}}
\dashline{20}(5,0)(100,0)
\end{picture}}

\newcommand{\ytwoDs}{\begin{picture}(\xmax,\ymax)(-\xx,\yy)
\setlength{\unitlength}{.1mm} \put(0,100){\circle*{10}}
\put(100,100){\circle*{10}} \put(100,0){\circle*{10}}
\put(0,0){\circle{10}} \put(0,100){\line(100,0){100}}
\put(0,100){\line(0,-100){95}} \put(100,100){\line(0,-100){95}}
\put(100,100){\line(-1,-1){95}} \put(0,100){\line(1,-1){95}}
\put(5,0){\line(100,0){100}}
\end{picture}}

\newcommand{\ytwoDss}{\begin{picture}(\xmax,\ymax)(-\xx,\yy)
\setlength{\unitlength}{.1mm} \put(0,100){\circle*{10}}
\put(100,100){\circle*{10}} \put(100,0){\circle*{10}}
\put(0,0){\circle{10}} \put(0,100){\line(100,0){100}}
\put(0,100){\line(0,-100){95}}
\put(100,100){\line(0,-100){95}}
\put(100,100){\line(-1,-1){95}} \put(0,100){\line(1,-1){95}}
\end{picture}}

\newcommand{\ytwoCd}{\begin{picture}(\xmax,\ymax)(-\xx,\yy)
\setlength{\unitlength}{.1mm} \put(0,100){\circle*{10}}
\put(100,100){\circle*{10}} \put(100,0){\circle*{10}}
\put(0,0){\circle{10}} \put(0,100){\line(100,0){95}}
\put(0,100){\line(0,-100){95}} \put(100,0){\line(0,100){95}}
\put(100,0){\line(-100,0){95}}
\dashline{20}(5,5)(100,100)
\end{picture}}

\newcommand{\ytwoAud}{\begin{picture}(\xmax,\ymax)(-\xx,\yy)
\setlength{\unitlength}{.1mm} \put(0,0){\circle*{10}}\put(0,100){\circle*{10}}
\put(100,100){\circle*{10}} \put(100,0){\circle*{10}}
 \put(0,100){\line(100,0){100}}
\put(0,100){\line(0,-100){95}} \put(100,100){\line(0,-100){95}}
\dashline{20}(5,0)(100,0)
\end{picture}}

\newcommand{\ytwoAus}{\begin{picture}(\xmax,\ymax)(-\xx,\yy)
\setlength{\unitlength}{.1mm} \put(0,0){\circle*{10}}\put(0,100){\circle*{10}}
\put(100,100){\circle*{10}} \put(100,0){\circle*{10}}
 \put(0,100){\line(100,0){100}}
\put(0,100){\line(0,-100){95}} \put(100,100){\line(0,-100){95}}
\put(0,0){\line(1,0){100}}
\end{picture}}

\newcommand{\ytwoBud}{\begin{picture}(\xmax,\ymax)(-\xx,\yy)
\setlength{\unitlength}{.1mm} \put(0,0){\circle*{10}}\put(0,100){\circle*{10}}
\put(100,100){\circle*{10}} \put(100,0){\circle*{10}}
 \put(0,100){\line(100,0){100}}
\put(0,100){\line(0,-100){95}} \put(100,100){\line(0,-100){95}}
\put(100,100){\line(-1,-1){100}}
\dashline{20}(5,0)(100,0)
\end{picture}}

\newcommand{\ytwoBus}{\begin{picture}(\xmax,\ymax)(-\xx,\yy)
\setlength{\unitlength}{.1mm} \put(0,0){\circle*{10}}\put(0,100){\circle*{10}}
\put(100,100){\circle*{10}} \put(100,0){\circle*{10}}
 \put(0,100){\line(100,0){100}}
\put(0,100){\line(0,-100){95}} \put(100,100){\line(0,-100){95}}
\put(100,100){\line(-1,-1){100}}
\put(0,0){\line(1,0){100}}
\end{picture}}

\newcommand{\ytwoCud}{\begin{picture}(\xmax,\ymax)(-\xx,\yy)
\setlength{\unitlength}{.1mm} \put(0,100){\circle*{10}}
\put(100,100){\circle*{10}} \put(100,0){\circle*{10}}
\put(0,0){\circle*{10}} \put(0,100){\line(100,0){95}}
\put(0,100){\line(0,-100){95}} \put(100,0){\line(0,100){95}}
\put(100,0){\line(-100,0){95}}
\dashline{20}(5,5)(100,100)
\end{picture}}

\newcommand{\ytwoDud}{\begin{picture}(\xmax,\ymax)(-\xx,\yy)
\setlength{\unitlength}{.1mm} \put(0,100){\circle*{10}}
\put(100,100){\circle*{10}} \put(100,0){\circle*{10}}
\put(0,0){\circle*{10}} \put(0,100){\line(100,0){100}}
\put(0,100){\line(0,-100){95}} \put(100,100){\line(0,-100){95}}
\put(100,100){\line(-1,-1){95}} \put(0,100){\line(1,-1){95}}
\dashline{20}(5,0)(100,0)
\end{picture}}

\newcommand{\ytwoDus}{\begin{picture}(\xmax,\ymax)(-\xx,\yy)
\setlength{\unitlength}{.1mm} \put(0,100){\circle*{10}}
\put(100,100){\circle*{10}} \put(100,0){\circle*{10}}
\put(0,0){\circle*{10}} \put(0,100){\line(100,0){100}}
\put(0,100){\line(0,-100){95}} \put(100,100){\line(0,-100){95}}
\put(100,100){\line(-1,-1){95}} \put(0,100){\line(1,-1){95}}
\put(0,0){\line(100,0){100}}
\end{picture}}

\newcommand{\ytwoAdhv}{\begin{picture}(\xmax,\ymax)(-\xx,\yy)
\setlength{\unitlength}{.1mm} \put(0,0){\circle*{10}}\put(0,100){\circle*{10}}
\put(100,100){\circle*{10}} \put(100,0){\circle*{10}}
 \put(0,100){\line(100,0){100}}
\put(0,100){\line(0,-100){95}} \put(100,100){\line(0,-100){95}}
\dashline{20}(5,0)(100,0)
\put(50,0){\vector(1,0){10}}
\put(100,50){\vector(0,1){10}}
\end{picture}}

\newcommand{\ytwoAdhh}{\begin{picture}(\xmax,\ymax)(-\xx,\yy)
\setlength{\unitlength}{.1mm} \put(0,0){\circle*{10}}\put(0,100){\circle*{10}}
\put(100,100){\circle*{10}} \put(100,0){\circle*{10}}
 \put(0,100){\line(100,0){100}}
\put(0,100){\line(0,-100){95}} \put(100,100){\line(0,-100){95}}
\dashline{20}(5,0)(100,0)
\put(50,0){\vector(1,0){10}}
\put(50,100){\vector(1,0){10}}
\end{picture}}

\newcommand{\ytwoBdhv}{\begin{picture}(\xmax,\ymax)(-\xx,\yy)
\setlength{\unitlength}{.1mm} \put(0,0){\circle*{10}}\put(0,100){\circle*{10}}
\put(100,100){\circle*{10}} \put(100,0){\circle*{10}}
 \put(0,100){\line(100,0){100}}
\put(0,100){\line(0,-100){95}} \put(100,100){\line(0,-100){95}}
\put(100,100){\line(-1,-1){100}}
\dashline{20}(5,0)(100,0)
\put(50,0){\vector(1,0){10}}
\put(100,50){\vector(0,1){10}}
\end{picture}}

\newcommand{\ytwoBdhd}{\begin{picture}(\xmax,\ymax)(-\xx,\yy)
\setlength{\unitlength}{.1mm} \put(0,0){\circle*{10}}\put(0,100){\circle*{10}}
\put(100,100){\circle*{10}} \put(100,0){\circle*{10}}
 \put(0,100){\line(100,0){100}}
\put(0,100){\line(0,-100){95}} \put(100,100){\line(0,-100){95}}
\put(100,100){\line(-1,-1){100}}
\dashline{20}(5,0)(100,0)
\put(50,0){\vector(1,0){10}}
\put(50,50){\vector(1,1){10}}
\end{picture}}

\newcommand{\ytwoDdhv}{\begin{picture}(\xmax,\ymax)(-\xx,\yy)
\setlength{\unitlength}{.1mm} \put(0,0){\circle*{10}}\put(0,100){\circle*{10}}
\put(100,100){\circle*{10}} \put(100,0){\circle*{10}}
 \put(0,100){\line(100,0){100}}
\put(0,100){\line(0,-100){95}} \put(100,100){\line(0,-100){95}}
\put(100,100){\line(-1,-1){100}}
\put(0,100){\line(1,-1){100}}
\dashline{20}(5,0)(100,0)
\put(50,0){\vector(1,0){10}}
\put(100,50){\vector(0,1){10}}
\end{picture}}

\begin{document}

\title{Simple relationship between the virial-route hypernetted-chain  and the compressibility-route Percus--Yevick  values of the fourth virial coefficient}

\author{Andr\'es Santos}
\email{andres@unex.es}
\homepage{http://www.unex.es/eweb/fisteor/andres/}
\author{Gema Manzano}
\affiliation{Departamento
de F\'{\i}sica, Universidad de Extremadura, E-06071 Badajoz, Spain}
\begin{abstract}
As is well known, approximate integral equations for liquids, such as the  hypernetted chain (HNC) and Percus--Yevick (PY) theories, are in general thermodynamically inconsistent in the sense that the macroscopic properties obtained from the spatial correlation functions depend on the route followed. In particular, the values of the fourth virial coefficient $B_4$ predicted by the HNC and PY approximations via the virial route differ from those obtained via the compressibility route. Despite this, it is shown in this paper that the value of $B_4$ obtained from the virial route in the HNC theory is exactly three halves the value obtained from the compressibility route in the PY theory, irrespective of the interaction potential (whether  isotropic or not), the number of components, and the dimensionality of the system. This simple relationship is confirmed in one-component systems by analytical results for the one-dimensional penetrable-square-well model and the three-dimensional penetrable-sphere model, as well as by numerical results for the one-dimensional Lennard--Jones model, the one-dimensional Gaussian core model, and the three-dimensional square-well model.

\end{abstract}
\maketitle

\maketitle

\section{Introduction and statement of the problem}
\label{sec2}

According to equilibrium statistical mechanics,\cite{B74,BH76,HM06} all the relevant structural and thermodynamic properties of  a one-component fluid made of particles interacting via a pairwise potential $\phi(\mathbf{r})=\phi(-\mathbf{r})$ are contained in the pair correlation function $g(\mathbf{r};\rho,\beta)=g(-\mathbf{r};\rho,\beta)$, where $\rho$ is the number density and $\beta=1/k_BT$ is the inverse temperature. In particular, the thermodynamic quantities of the fluid can be obtained from $g(\mathbf{r};\rho,\beta)$ through different routes. The most common ones are the virial route,
\beq
Z(\rho,\beta)\equiv\frac{\beta p}{\rho}
=1+\frac{\rho}{2d}\int \dd \mathbf{\rr}\,
y(\rr;\rho,\beta)\rr\cdot\nabla f(\rr;\beta),
\label{1}
\eeq
the compressibility route,
\beqa
\chi(\rho,\beta)&\equiv&\left(\beta\frac{\partial p}{\partial \rho}\right)^{-1}
\nn
&=& 1+\rho\int \dd \mathbf{\rr}\,
\left\{\left[f(\rr;\beta)+1\right]y(\rr;\rho,\beta)-1\right\},
\label{2}
\eeqa
and the energy route
\beq
u(\rho,\beta)= \frac{d}{2\beta}-\frac{\rho}{2}\int \dd
\mathbf{r}\, y(\rr;\rho,\beta)\frac{\partial}{\partial\beta}f(\rr;\beta) .
\label{3}
\eeq
In Eqs.\ (\ref{1})--(\ref{3}), $p$ is the pressure, $Z$ is the compressibility factor, $d$ is the dimensionality of the system,  $\chi$ is the isothermal susceptibility, $u$ is the internal energy per particle,
$f(\rr)\equiv e^{-\beta \phi(\rr)}-1$ is the Mayer function, and
$y(\rr)\equiv \exp[\beta\phi(\rr)]g(\rr)$ is the so-called cavity function.
Note that in Eq.\ \eqref{2} the integrand $[f(\rr)+1]y(\rr)-1 =
g(\rr) - 1 \equiv h(\rr)$ is the total correlation function. It is re-
lated to the direct correlation function $c(\rr)$ through the
Ornstein--Zernike (OZ) relation\cite{HM06}
\beq
h(\rr) = c(\rr) + \rho\int\dd\rr'\, h(\rr-\rr')c(\rr').
\label{OZ}
\eeq
In terms of $c(\rr)$, the compressibility equation of state \eqref{2} can
be rewritten as
\beq
\chi^{-1}=1-\rho\widetilde{c},
\label{chi1}
\eeq
where
\beq
\widetilde{c}=\int\dd\rr\, c(\rr)
\label{FT}
\eeq
is the spatial integral of $c(\rr)$ or, equivalently, its Fourier transform  at zero wavenumber.

Thermodynamic relations establish that
\beq
\chi^{-1}(\rho,\beta)=\frac{\partial}{\partial\rho}\left[\rho
Z(\rho,\beta)\right],
\label{4}
\eeq
\beq
\rho\frac{\partial}{\partial\rho}u(\rho,\beta)=\frac{\partial}{\partial
\beta}{Z(\rho,\beta)}.
\label{5}
\eeq
Of course, the \emph{exact} correlation function $g^\ex(\rr;\rho,\beta)$ yields functions $Z(\rho,\beta)$, $\chi(\rho,\beta)$, and $u(\rho,\beta)$ satisfying Eqs.\ \eqref{4} and \eqref{5}.
On the other hand, except in one-dimensional problems restricted to nearest-neighbor interactions, the function $g^\ex(\rr;\rho,\beta)$ is not known and so one has to resort to approximations. The price to be paid is that, in general, an approximate function $g^\app(\rr;\rho,\beta)$ violates Eqs.\ \eqref{4} and \eqref{5} and so it is thermodynamically \emph{inconsistent}.
There exist, however, a small number of approximations in which the energy route is consistent with the virial route, i.e., Eq.\ \eqref{5} is verified. This class of approximations includes the hypernetted chain\cite{M60} (HNC)  and the linearized Debye--H\"uckel\cite{SFG09} theories for arbitrary potentials, the mean-spherical approximation for soft-potentials,\cite{S07} and  the hard-sphere limit of the square-shoulder potential for any approximation.\cite{S05,S06}
On the other hand, even in those cases the virial-compressibility consistency condition, Eq.\ \eqref{4}, is not satisfied.
In fact, some liquid state
theories include one or more adjustable state-dependent parameters
which are tuned to achieve thermodynamic consistency between two or
more routes. This is the case, for instance, of the modified
HNC closure,\cite{RA79,CPE97} the Rogers--Young
closure,\cite{RY84} the Zerah--Hansen closure,\cite{ZH85} the
self-consistent Ornstein--Zernike approximation,\cite{HS77} the
hierarchical reference theory,\cite{PR95} Lee's theory based on the
zero-separation theorems,\cite{L95} the generalized MSA,\cite{W73} or the rational-function
approximation.\cite{YS91}

The thermodynamic inconsistency problem manifests itself at the level of the virial coefficients $B_n(\beta)$. They are defined by the  series expansion
\beq
Z(\rho,\beta)=1+\sum_{n=1}^\infty B_{n+1}(\beta)\rho^n.
\label{8}
\eeq
Likewise, the series expansion of the isothermal susceptibility, the internal energy, and the cavity function in powers of density can be written as
\beq
\chi(\rho,\beta)=1+\sum_{n=1}^\infty
\chi_{n+1}(\beta)\rho^n,
\label{8c}
\eeq\beq
 u(\rho,\beta)=\frac{d}{2\beta}+\sum_{n=1}^\infty
u_{n+1}(\beta)\rho^n,
\label{8e}
\eeq
\beq
y(\rr;\rho,\beta) = 1+\sum_{n=1}^\infty y_n(\rr;\beta)\rho^n.
\label{eq7}
\eeq
Inserting Eq.\ \eqref{eq7} into Eq.\ \eqref{1} one can express $B_n$ in terms of an integral involving $y_{n-2}$,
\beq
B_{n,\vi}(\beta)=\frac{1}{2d}\int \dd \mathbf{r}\,
y_{n-2}(\rr;\beta)\rr\cdot\nabla f(\rr;\beta),\quad n\geq 2,
\label{9}
\eeq
where we use the subscript ``v'' to emphasize that Eq.\ \eqref{9} gives the $n$th virial coefficient through the \emph{virial} route.
Similarly, insertion of Eq.\ \eqref{eq7} into Eqs.\ \eqref{2} and \eqref{3} yields
\beq
\chi_2(\beta)= \int \dd \mathbf{r}\,  f(\rr;\beta),
\label{10a}
\eeq
\beq
\chi_n(\beta)=\int \dd \mathbf{r}\,
\left[f(\rr;\beta)+1\right]y_{n-2}(\rr;\beta),\quad n\geq 3,
\label{10}
\eeq
\beq
u_n(\beta)= -\frac{1}{2}\int \dd \mathbf{r}\,
y_{n-2}(\rr;\beta)\frac{\partial}{\partial\beta}f(\rr;\beta),\quad n\geq 2 .
\label{11}
\eeq
Use of the thermodynamic relations \eqref{4} and \eqref{5} allows one to express the virial coefficients $B_n$ from the coefficients $\chi_n$ and $u_n$, respectively, as
\beq
B_{n,\co}(\beta)=-\frac{1}{n}\sum_{m=1}^{n-1}mB_{m,\co}(\beta)\chi_{n+1-m}(\beta)
,\quad n\geq 2,
\label{11b}
\eeq
\beq
u_n(\beta)=\frac{1}{n-1}\frac{\partial}{\partial\beta}B_{n,\en}(\beta),\quad n\geq 2.
\label{12}
\eeq
Here the subscripts ``c'' and ``e'' denote virial coefficients obtained via the \emph{compressibility} and \emph{energy}  routes, respectively. Obviously, the exact functions $y_n^\ex$ give consistent virial coefficients, i.e., $B_{n,\vi}^{\ex}=B_{n,\co}^{\ex}=B_{n,\en}^{\ex}$. As said above, the HNC theory provides consistent thermodynamic properties through the energy and virial routes only, i.e.,  $B_{n,\vi}^{\hnc}=B_{n,\en}^{\hnc}\neq B_{n,\co}^{\hnc}$. Except for a few cases,\cite{SFG09,S07,S05,S06} an approximate theory generally predicts  different sets of thermodynamic quantities from the three routes. This is the case of the well-known Percus--Yevick (PY) theory,\cite{B74,BH76,HM06} i.e., $B_{n,\vi}^{\py}\neq B_{n,\en}^{\py}\neq B_{n,\co}^{\py}$.

Most of the liquid state theories, including HNC and PY, give the exact cavity function to first order in density and hence provide the exact second and third virial coefficients. Therefore, the fourth virial coefficient is usually the earliest one revealing the approximate nature of the theory. In principle, the two HNC coefficients ($B_{4,\vi}^{\hnc}=B_{4,\en}^{\hnc}$ and $B_{4,\co}^{\hnc}$) and the three PY coefficients ($B_{4,\vi}^{\py}$, $B_{4,\en}^{\py}$, and $B_{4,\co}^{\py}$) are unrelated. On the other hand, recent analytical evaluations of the fourth virial coefficient for the three-dimensional penetrable-sphere potential\cite{SM07} and for the one-dimensional penetrable-square-well potential\cite{SFG08} show a remarkably simple relationship between $B_{4,\vi}^{\hnc}$ and $B_{4,\co}^{\py}$, namely
\beq
B_{4,\vi}^{\hnc}(\beta)=\frac{3}{2}B_{4,\co}^{\py}(\beta),
\label{1.1}
\eeq
where, according to Eq.\ \eqref{11b},
\beq
B_{4,\co}=\frac{1}{4}\left(2\chi_2\chi_3-\chi_2^3-\chi_4\right).
\label{1.2}
\eeq
The two potentials considered in Refs.\ \onlinecite{SM07,SFG08} have  two common features:  they are (i) bounded and (ii) stepwise constant. Thus, it might be the case that the relation \eqref{1.1} is closely connected to one or both features, not being generally valid. The aim of this paper is to show that this is not the case by proving Eq.\ \eqref{1.1} for any dimensionality $d$ and any potential $\phi(\rr)$. Furthermore, we will show that Eq.\ \eqref{1.1} can be extended to fluid mixtures with any number of components.

This paper is organized as follows. The diagrammatic representation of the cavity function to second order in density and of the fourth virial coefficient is presented in Sec.\ \ref{sec3}. This is followed by the  mathematical proof of Eq.\ \eqref{1.1}, where use is made of identities derived in the Appendix. Section \ref{sec4} provides a few examples where Eq.\ \eqref{1.1} is numerically verified. Next, it is shown in Sec.\ \ref{sec4b} that the results remain valid in the more general case of a multicomponent fluid. Finally, the results are discussed in Sec.\ \ref{sec5}.

\section{Proof of the relationship $B_{4,\vi}^{\hnc}=\frac{3}{2}B_{4,\co}^{\py}$}
\label{sec3}

Statistical-mechanical methods allow one to express the functions $y_n(\rr;\beta)$ as sums of $n$-particle multiple integrals of products of Mayer functions. These integrals are conveniently represented by diagrams.\cite{B74,BH76,HM06} In particular,
\beq
y_1(\rr;\beta)=\yone,
\label{2.3}
\eeq
\beqa
y_2^\ex(\rr;\beta)&=&\frac{1}{2}\Bigg(2\ytwoA+4\ytwoB\nn
&&+\ytwoC+\ytwoD\Bigg)
.
\label{2.4}
\eeqa
The open circles represent two fixed \textit{root} points separated by a
vector $\rr_{12}=\rr$, the filled circles represent \textit{field} points to
be integrated out, and each bond represents a Mayer function. For instance,
\beq
\yone=\int \dd3 \,f_{13}f_{23},
\label{2.6}
\eeq
\beq
\ytwoB=
\int \dd3\int \dd4 \, f_{13}f_{34}
f_{24}f_{14},
\label{2.7}
\eeq
where $\dd i\equiv \dd\rr_i$ and $f_{ij}=f_{ji}\equiv f(\mathbf{r}_i-\mathbf{r}_j;\beta)$.
The factors $2$ and $4$ affecting the first and second diagrams of Eq. \eqref{2.4}, respectively, reflect the degeneracies of those diagrams with respect to the exchange of field points or of root points. As will be seen in Sec.\ \ref{sec4b}, these degeneracies are broken down in the multicomponent case.

Both the HNC and the PY approximations are consistent with Eq.\ \eqref{2.3} but not with Eq.\ \eqref{2.4}. In the case of HNC, the last diagram on the right-hand side of Eq.\ \eqref{2.4} is neglected, while PY neglects the  last two diagrams. More in general, imagine an approximation that includes those  last two diagrams but with weights $\lambda_1$ and $\lambda_2$, respectively:
\beqa
y_2^{(\lambda_1,\lambda_2)}(\rr;\beta)&=&\frac{1}{2}\Bigg(2\ytwoA+4\ytwoB\nn
&&+\lambda_1\ytwoC+\lambda_2\ytwoD\Bigg).
\label{18}
\eeqa
The class of functions $y_2^{(\lambda_1,\lambda_2)}(\rr;\beta)$ includes the exact, HNC, and PY functions as special cases:
\beqa
y_2^\ex(\rr;\beta)&=&y_2^{(1,1)}(\rr;\beta),\nn
y_2^\hnc(\rr;\beta)&=&y_2^{(1,0)}(\rr;\beta),\\
y_2^\py(\rr;\beta)&=&y_2^{(0,0)}(\rr;\beta)\nonumber.
\label{17}
\eeqa

Substitution of Eq.\ \eqref{18} into Eq.\ \eqref{9}  for $n=4$ yields
\beqa
B_{4,\vi}^{(\lambda_1,\lambda_2)}(\beta)
&=&\frac{1}{4d}\Bigg(2\ytwoAd+4\ytwoBd\nn
&&+\lambda_1\ytwoCd+\lambda_2\ytwoDd\Bigg),
\label{19}
\eeqa
where here one of the root points has become a field point and a dashed bond represents a term of the form $\rr\cdot \nabla f(\rr)$. For instance,
\beq
\ytwoBd=\int \dd2\int\dd3\int \dd4 \, f_{13}f_{34}
f_{24}f_{14}\rr_{12}\cdot\nabla_{12} f_{12}.
\label{2.8}
\eeq
Note that the one-root diagrams do not depend on the location of the root, so they can be expressed as zero-root diagrams divided by the $d$-dimensional volume $V$. For example,
\beq
\ytwoBd=\ytwoBdd=\frac{1}{V}\ytwoBud.
\label{2.9}
\eeq

In the case of the isothermal susceptibility coefficients, Eqs.\ \eqref{10a}, \eqref{10}, \eqref{2.3}, and \eqref{18} give
\beq
\chi_2=\chitwo,
\label{21}
\eeq
\beq
\chi_3=\yoneA+\yones,
\label{22}
\eeq
\beqa
\chi_{4}^{(\lambda_1,\lambda_2)}(\beta)
&=&\ytwoAA+\frac{2+\lambda_1}{2}\ytwoAs+2\ytwoBB\nn
&&+\frac{4+\lambda_1+\lambda_2}{2}\ytwoBs+\frac{\lambda_2}{2}\ytwoDs.
\label{20}
\eeqa
In Eq.\ \eqref{20} we have taken into account that
\beq
\ytwoDss=\ytwoBss=\ytwoBs.
\label{2.10}
\eeq
Equations \eqref{1.2} and \eqref{21}--\eqref{20} readily give
\beqa
B_{4,\co}^{(\lambda_1,\lambda_2)}(\beta)
&=&-\frac{2+\lambda_1}{8}\ytwoAs-\frac{4+\lambda_1+\lambda_2}{8}\ytwoBs\nn
&&-\frac{\lambda_2}{8}\ytwoDs,
\label{24}
\eeqa
where use has been made of the properties
\beq
\chi_2\chi_3=\ytwoAA+\ytwoBB,\quad \chi_2^3=\ytwoAA.
\label{23}
\eeq
Equation \eqref{24} can also be obtained from Eq.\ \eqref{chi1}. This is the path followed in Sec.\ \ref{sec4b} for the multicomponent case.

Now, in order to connect Eqs.\ \eqref{19} and \eqref{24}, we need to make use of the following properties\cite{BH76} (see the Appendix for a derivation):
\beq
\ytwoAd=-\frac{3d}{4}\ytwoAs,
\label{25}
\eeq
\beq
\ytwoBd+\frac{1}{4}\ytwoCd=-\frac{3d}{4}\ytwoBs,
\label{26}
\eeq
\beq
\ytwoDd=-\frac{d}{2}\ytwoDs.
\label{27}
\eeq
Therefore, Eq.\ \eqref{19} can be rewritten as
\beqa
B_{4,\vi}^{(\lambda_1,\lambda_2)}(\beta)&=&-\frac{3}{8}\ytwoAs-\frac{3}{4}\ytwoBs-\frac{\lambda_2}{8}\ytwoDs\nn
&&+\frac{\lambda_1-1}{4d}\ytwoCd.
\label{28}
\eeqa
Comparison between Eqs.\ \eqref{24} and \eqref{28} shows that
\beq
B_{4,\vi}^{(1,\lambda_2)}(\beta)=\frac{3}{2+\lambda_1}B_{4,\co}^{(\lambda_1,\lambda_1)}(\beta),
\label{29}
\eeq
provided that
\beq
\lambda_2=\frac{3\lambda_1}{2+\lambda_1}.
\label{30}
\eeq
This is the main result of this paper. If $\lambda_1=1$ then $\lambda_2=1$ and we recover the exact consistency condition: $B_{4,\vi}^\ex=B_{4,\co}^\ex$. On the other hand, the choice $\lambda_1=0$ yields the sought result, Eq.\ \eqref{1.1}.

\begin{figure}[h]
  \includegraphics[width=0.95\columnwidth]{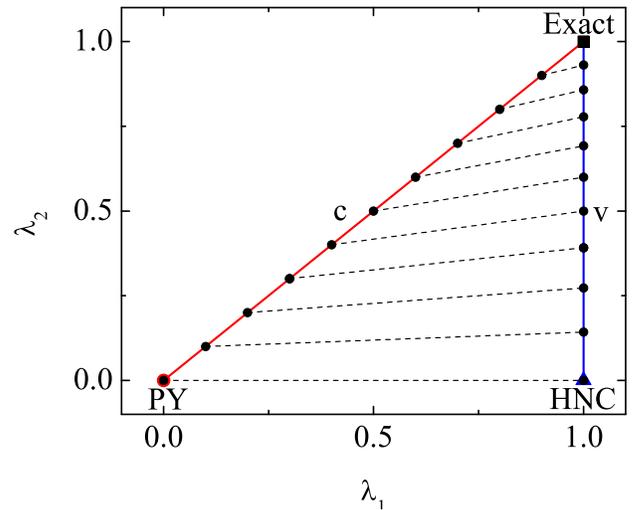}
\caption{ The diagonal (labeled c) and vertical (labeled v) lines represent the  $(\lambda_1,\lambda_1)$-subclass and the $(1,\lambda_2)$-subclass of approximations, respectively. Both lines intersect in the exact case $\lambda_1=\lambda_2=1$, while the other ends correspond to the PY and HNC approximations, respectively. The virial-route value of the fourth virial coefficient in any $(1,\lambda_2)$-subclass approximation is related to the compressibility-route value in a given  $(\lambda_1,\lambda_1)$-subclass approximation. This is represented by the dots connected by dashed lines.}\label{diagram}
\end{figure}

It is quite apparent that the Eq.\ \eqref{29} is, from a mathematical point of view, more general than \eqref{1.1}.  Let us define the $(\lambda_1,\lambda_1)$-subclass of approximations as the one compatible with Eq.\ \eqref{18} with $0\leq\lambda_1=\lambda_2\leq 1$. Analogously, we define the $(1,\lambda_2)$-subclass of approximations as the one with $\lambda_1=1$ and $0\leq\lambda_2\leq 1$. Obviously, any approximation retaining the exact $y_2(r)$ (i.e., $\lambda_1=\lambda_2=1$) belongs to both subclasses. On the other hand, the PY and HNC theories are members of the $(\lambda_1,\lambda_1)$-subclass and $(1,\lambda_2)$-subclass, respectively. Equations \eqref{29} and \eqref{30} then state that, for any $(\lambda_1,\lambda_1)$-subclass approximation, there exists a specific $(1,\lambda_2)$-subclass approximation, such that the compressibility and virial values, respectively, of $B_4$ are proportional each other.  The connection between both subclasses is schematically illustrated in Fig.\ \ref{diagram}.
Interestingly, the largest deviation of the proportionality factor from 1 occurs at $\lambda_1=0$, i.e., in the case of the PY and HNC pair.

\section{Numerical examples}
\label{sec4}

\begin{figure}[h]
 \includegraphics[width=0.95\columnwidth]{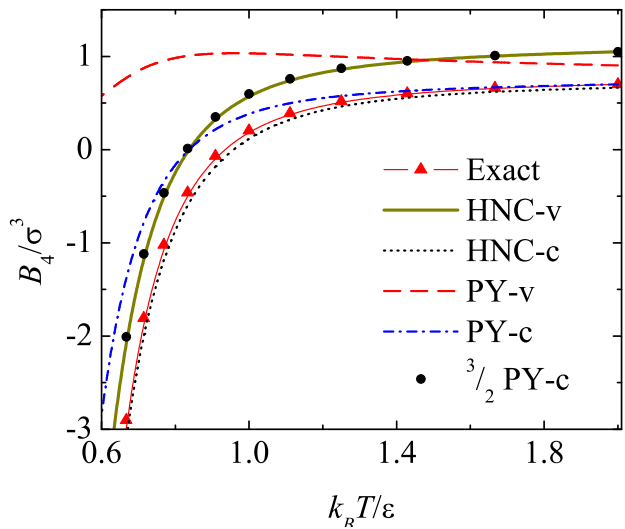}
\caption{ Temperature dependence of the fourth virial coefficient for the one-dimensional LJ potential. The thin line with triangles represent the exact results, while $B_{4,\vi}^\hnc$, $B_{4,\co}^\hnc$, $B_{4,\vi}^\py$, and $B_{4,\co}^\py$ are represented by the thick solid, dotted, dashed, and dash-dotted lines, respectively. The circles represent the values of $\frac{3}{2} B_{4,\co}^\py$, which fall on the $B_{4,\vi}^\hnc$ curve.}\label{B4LJ}
\end{figure}

\begin{figure}[h]
  \includegraphics[width=0.95\columnwidth]{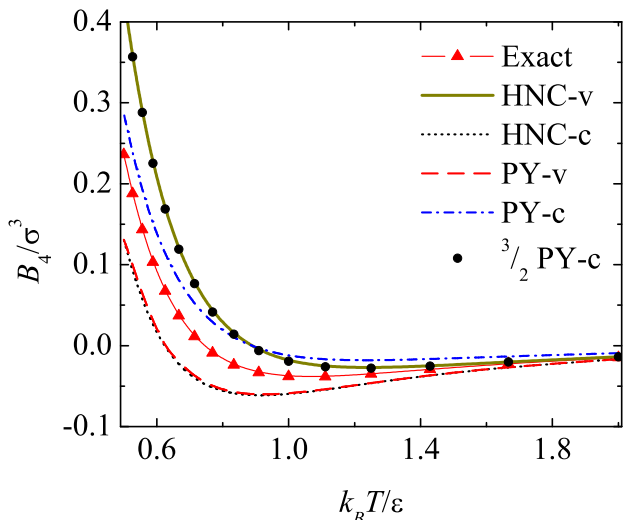}
\caption{ Same as in Fig.\ \protect\ref{B4LJ}, but for the one-dimensional Gaussian potential. Note that the  $B_{4,\co}^\hnc$ and $B_{4,\vi}^\py$ curves are hardly distinguishable in this case.}\label{B4G}
\end{figure}

\begin{figure}[h]
  \includegraphics[width=0.95\columnwidth]{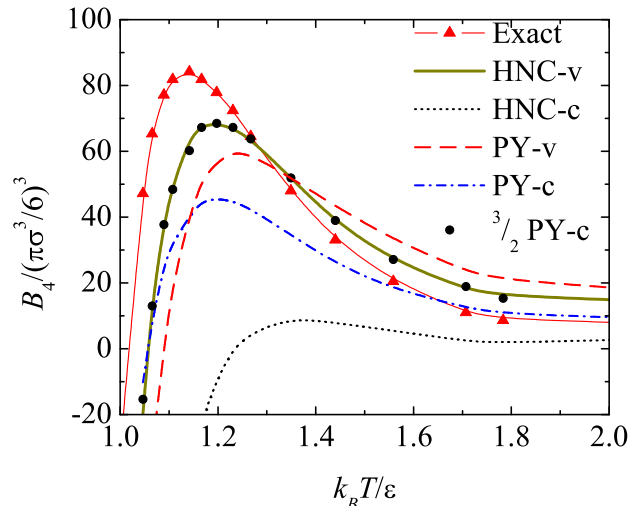}
\caption{ Same as in Fig.\ \protect\ref{B4LJ}, but for the three-dimensional SW potential with $\lambda=1.5$. The data have been extracted from Fig.\ 10 of Ref.\ \protect\onlinecite{BH67}.}\label{B4SW}
\end{figure}

As said in Sec.\ \ref{sec2}, the simple relationship \eqref{1.1} derived in the preceding section is confirmed by analytical results for the three-dimensional penetrable-sphere model\cite{SM07} and for the one-dimensional penetrable-square-well potential.\cite{SFG08}
The analytical derivation of the exact, HNC, and PY expressions of the fourth virial coefficient was possible in Refs.\ \onlinecite{SM07,SFG08} thanks to the simplicity of the interaction models (stepwise constant). In general, however, the computation of $B_4$ is a complicated task that requires numerical methods, especially in the case of continuous potentials.

As a numerical test of Eq.\ \eqref{1.1}, we have numerically evaluated $B_{4}^\ex$, $B_{4,\vi}^\hnc$, $B_{4,\co}^\hnc$, $B_{4,\vi}^\py$, and $B_{4,\co}^\py$ for two one-dimensional interaction models. The first one is the conventional Lennard--Jones (LJ) potential:
\beq
\phi(\rr)=4\epsilon\left[\left(\frac{\sigma}{r}\right)^{12}-\left(\frac{\sigma}{r}\right)^{6}\right].
\label{4.1}
\eeq
This is an unbounded potential, repulsive for $r<2^{1/6}\sigma$ and attractive for $r>2^{1/6}\sigma$. As a second example, we have considered the bounded, purely repulsive Gaussian core model,\cite{SS97}
\beq
\phi(\rr)=\epsilon \exp\left[-(r/\sigma)^2\right].
\label{4.2}
\eeq
The results are displayed in Figs.\ \ref{B4LJ} and \ref{B4G}, respectively. In the case of the LJ potential, Fig.\ \ref{B4LJ} shows that the best general agreement with the exact results are provided by $B_{4,\co}^\hnc$ and the worst by $B_{4,\vi}^\py$. As the temperature increases, $B_{4,\co}^\py$ eventually becomes quite accurate. What is more important from the point of view of this work is that the numerical values of $\frac{3}{2}B_{4,\co}^\py$ are practically indistinguishable from those of $B_{4,\vi}^\hnc$, in agreement with Eq.\ \eqref{1.1}.

Regarding the Gaussian core potential, Fig.\ \ref{B4G} shows that $B_{4,\co}^\py$ provides the best description for $k_BT/\epsilon\lesssim 0.9$, while $B_{4,\vi}^\hnc$ does for $k_BT/\epsilon\gtrsim 0.9$.
It is also noteworthy that $B_{4,\co}^\hnc\simeq B_{4,\vi}^\py$ for the whole temperature range shown. Again, the numerical results confirm Eq.\ \eqref{1.1}.

As a final test, we have turned our attention to the pioneering numerical evaluation by Barker and Henderson\cite{BH67} of the fourth virial coefficient (they also considered the fifth one) for the three-dimensional square-well (SW) potential
\beq
\phi(\rr)=\begin{cases}
\infty,&r<\sigma,\\
-\epsilon,&\sigma<r<\lambda\sigma,\\
0,&r>\lambda\sigma.
\end{cases}
\label{4.3}
\eeq
The results for a potential range $\lambda=1.5$ are displayed in Fig.\ \ref{B4SW}. In this case, $B_{4,\vi}^\hnc$ and $B_{4,\co}^\py$ give the most accurate results for $1\lesssim k_BT/\epsilon\lesssim 1.5$ and $k_BT/\epsilon\gtrsim 1.5$, respectively. Moreover, both approximations are observed to be consistent with Eq.\ \eqref{1.1} within unavoidable numerical uncertainties.

\section{Extension to fluid mixtures\label{sec4b}}
So far, in Secs.\ \ref{sec2}--\ref{sec4} we have assumed a one-component
fluid. However, the main result of this paper, Eq.\ \eqref{1.1},
can be extended to the case of a mixture, as shown below.

Let us consider a multicomponent fluid made of an arbitrary number of species with mole fractions $x_i$, at density
$\rho$ and inverse temperature $\beta$. The interaction potential
between a particle of species $i$ and a particle of species
$j$ is $\phi_{ij}(\rr) = \phi_{ji}(-\rr)$. The pair correlation and cavity functions are $g_{ij}(\rr;\rho;\beta;\{x_k\})$ and $y_{ij}(\rr;\rho;\beta;\{x_k\}) =
\exp [\beta\phi_{ij}(\rr)] g_{ij}(\rr;\rho;\beta;\{x_k\})$, respectively. The multicomponent version of Eq.\ \eqref{1} is simply
\beq
Z = 1 +\frac{\rho}{2d}\sum_{i,j}x_ix_j\int
\dd\rr\, y_{ij}(\rr)\rr \cdot \nabla f_{ij}(\rr).
\label{5.1}
\eeq
The compressibility equation of state  for the multicomponent case is more conveniently expressed as a generalization of Eq.\ \eqref{chi1} rather than of Eq.\ \eqref{2}. It is given by
\cite{BH06}
\beq
\chi^{-1}=1-\rho \sum_{i,j}x_ix_j\widetilde{c}_{ij},
\label{5.2}
\eeq
where $\widetilde{c}_{ij}=\widetilde{c}_{ji}$ are spatial integrals of the direct correlation functions $c_{ij}(\rr)$. The latter are defined in
terms of the total correlation functions $h_{ij}(\rr)\equiv g_{ij}(\rr)-1$
via the OZ relation
\beq
h_{ij}(\rr)=c_{ij}(\rr)+\rho\sum_k x_k \int\dd\rr'\, h_{ik}(\rr-\rr')c_{kj}(\rr').
\label{5.3}
\eeq
This equation yields
\beq
\widetilde{h}_{ij}=\widetilde{c}_{ij}+\rho\sum_k x_k \widetilde{h}_{ik}\widetilde{c}_{kj},
\label{5.4}
\eeq
where again the tilde denotes spatial integration.

Let us now consider the virial expansion. The virial coefficients $B_n$ are still defined by Eq.\ \eqref{8}, but now they can be expressed in terms of composition-independent coefficients. For instance,
\beq
B_{2}=\sum_{i,j} x_i x_j B_{ij},
\label{5.15_2}
\eeq
\beq
B_{3}=\sum_{i,j} x_i x_j x_k B_{ijk},
\label{5.15_3}
\eeq
\beq
B_{4}=\sum_{i,j,k,\ell} x_i x_j x_k x_\ell B_{ijk\ell}.
\label{5.15_4}
\eeq
As for the correlation functions, the generalization of Eq.\ \eqref{eq7} reads
\beq
y_{ij}(\rr)=1+\sum_{n=1}^\infty y_{ij;n}(\rr)\rho^n.
\label{5.5}
\eeq
A similar expansion can be carried out for $h_{ij}(\rr)$ and $c_{ij}(\rr)$. In particular,
\beq
\widetilde{h}_{ij;0}=\widetilde{f}_{ij},
\label{5.6}
\eeq
\beq
\widetilde{h}_{ij;n}=\widetilde{y}_{ij;n}+\widetilde{fy}_{ij;n},\quad n\geq 1,
\label{5.7}
\eeq
\beq
\widetilde{c}_{ij;0}=\widetilde{f}_{ij},
\label{5.8}
\eeq
\beq
\widetilde{c}_{ij;1}=\widetilde{y}_{ij;1}+\widetilde{fy}_{ij;1}-\sum_k x_k \widetilde{f}_{ik}\widetilde{f}_{kj},
\label{5.9}
\eeq
\beqa
\widetilde{c}_{ij;2}&=&\widetilde{y}_{ij;2}+\widetilde{fy}_{ij;2}-\sum_k x_k \widetilde{f}_{ik}\left(\widetilde{y}_{kj;1}+\widetilde{fy}_{kj;1}\right)\nn
&&-\sum_k x_k \widetilde{f}_{jk}\left(\widetilde{y}_{ki;1}+\widetilde{fy}_{ki;1}\right)+\sum_{k,\ell} x_k x_\ell \widetilde{f}_{ik}\widetilde{f}_{k\ell}\widetilde{f}_{\ell j}.\nn
\label{5.10}
\eeqa
In these equations, $\widetilde{fy}_{ij;n}$ denotes the spatial integral of $f_{ij}(\rr)y_{ij;n}(\rr)$

The coefficients $y_{ij;n}(\rr)$ in Eq.\ \eqref{5.5} are polynomials of degree $n$ in the mole fractions. In particular,
\beq
y_{ij;1}(\rr)=\sum_k x_k y_{ij|k}(\rr),
\label{5.11}
\eeq
\beq
y_{ij;2}(\rr)=\sum_{k,\ell} x_k  x_\ell y_{ij|k\ell}(\rr).
\label{5.12}
\eeq
Analogously,
\beq
c_{ij;2}(\rr)=\sum_{k,\ell} x_k  x_\ell c_{ij|k\ell}(\rr).
\label{5.13}
\eeq
{}From Eqs.\ \eqref{4} and \eqref{5.2} it follows that the fourth virial coefficient provided by the compressibility route is
\beq
B_{4,\co}=-\frac{1}{4}\sum_{i,j}x_i x_j \widetilde{c}_{ij;2},
\label{5.14}
\eeq
so that
\beqa
B_{ijk\ell,\co}&=&-\frac{1}{24}\left(\widetilde{c}_{ij;k\ell}+\widetilde{c}_{ik;j\ell}+\widetilde{c}_{i\ell;jk}+\widetilde{c}_{jk;i\ell}
\right.\nn
&&\left.+
\widetilde{c}_{j\ell;ik}+\widetilde{c}_{k\ell;ij}\right).
\label{5.16}
\eeqa

The function $y_{ij|k}(\rr)$ is represented by the same diagram as in Eq.\ \eqref{2.3}, except that the two root points must be labeled with $i$ and $j$ and the field point must be labeled with $k$. Consequently, $\widetilde{y}_{ij|k}= \widetilde{f}_{ik}\widetilde{f}_{kj}$. Thus, Eq.\ \eqref{5.10} yields
\beqa
\widetilde{c}_{ij;k\ell}&=&\widetilde{y}_{ij;k\ell}+\widetilde{fy}_{ij;k\ell}- \frac{1}{2}\left(\widetilde{f}_{ik}\widetilde{fy}_{kj;\ell}+ \widetilde{f}_{i\ell}\widetilde{fy}_{\ell j;k}\right.\nn
&&\left.+
\widetilde{f}_{jk}\widetilde{fy}_{ki;\ell}+ \widetilde{f}_{j\ell}\widetilde{fy}_{\ell i;k}\right)
- \widetilde{f}_{ik}\widetilde{f}_{k\ell}\widetilde{f}_{\ell j}.
\label{5.17}
\eeqa
As for the function $y_{ij|k\ell}(\rr)$, it is represented by the same diagrams as in Eq.\ \eqref{2.4}, except that (a) the root points are labeled with $i$ and $j$, while the field points are labeled with $k$ and $\ell$, and (b) the first and second diagrams on the right-hand side of Eq.\ \eqref{2.4}, with their corresponding factors, actually become 2 and 4 labeled diagrams, respectively. In other words, each diagram  is replaced by a symmetrized sum of topologically analogous diagrams, divided by its number, so that the symmetry property $y_{ij|k\ell}(\rr)=y_{ij|\ell k}(\rr)$ is preserved. We will refer to this process of generating the diagrams of the multicomponent case from those of the one-component case as \emph{symmetrization}.

Analogously to Eq.\ \eqref{18}, we can introduce the class of approximations $y_{ij|k\ell}^{(\lambda_1,\lambda_2)}(\rr)$. Therefore, Eqs.\ \eqref{5.1} and \eqref{5.15_4} imply that $B_{ijk\ell,\vi}^{(\lambda_1,\lambda_2)}$ is given by Eq.\ \eqref{19}, except that again each diagram is symmetrized to preserve the symmetry of $B_{ijk\ell}$ under any permutation of indices.
In the case of $B_{ijk\ell,\co}^{(\lambda_1,\lambda_2)}$, use of Eqs.\ \ref{5.16} and \eqref{5.17} shows that it is given by Eq.\ \eqref{24}, again with the symmetrization criterion.

The properties \eqref{25}--\eqref{27} do not necessarily hold for individual labeled diagrams, but they do when again each diagram is replaced by its symmetrized sum. As a consequence, one has
\beq
B_{ijk\ell,\vi}^{(1,\lambda_2)}=\frac{3}{2+\lambda_1}B_{ijk\ell,\co}^{(\lambda_1,\lambda_1)}\quad \lambda_2=\frac{3\lambda_1}{2+\lambda_1}.
\label{5.18}
\eeq
This generalizes Eq.\ \eqref{29}, and hence Eq.\ \eqref{1.1}, to the case of an arbitrary mixture.

\section{Discussion}
\label{sec5}
Most of the integral equation theories are consistent with the exact correlation functions to first order in density, Eq.\ \eqref{2.3}, so that they agree with the exact third virial coefficient. Therefore, the fourth virial coefficient is the earliest one that, not only differs from theory to theory, but even among different thermodynamic routes within the same theory. Thus, if one considers the two most  studied liquid state theories (HNC and PY) and the three most standard thermodynamic routes (virial, compressibility, and energy), there are in principle six alternative approximations for the fourth virial coefficient of a given interaction model. Actually, this number is reduced from six to five because the HNC theory belongs to the ``exclusive'' class of approximations that are thermodynamically consistent with respect to the energy and virial routes,\cite{M60,SFG09,S07} without being forced to do so.

The main aim of this paper has been to prove that the number of independent predictions given by the HNC and PY approximations further reduces from five to four because the fourth virial coefficient obtained from the virial route in the HNC theory is exactly three halves the value obtained from the compressibility route in the PY theory, for any interaction potential, any number of components,  and any dimensionality.
This result has been derived as a special case of a more general mathematical property described by Eqs.\ \eqref{29} and \eqref{30} and graphically sketched by Fig.\ \ref{diagram}.

It is interesting to remark that, as a simple corollary of Eq.\ \eqref{1.1},
the Boyle-like temperature $T_{B_4}$ at which $B_4=0$ is the same in the PY-c and HNC-v approximations. Since both predictions cross at $T=T_{B_4}$, none of them is closer to the exact value than the other one for the whole temperature range, as illustrated by Figs.\ \ref{B4LJ}--\ref{B4SW}.

One may reasonably wonder if a relation similar to Eq.\ \eqref{1.1} extends to higher-order virial coefficients. While we are not in conditions of ascertaining this possibility at this point, it is at least clear from Fig.\ 12 of Ref.\ \onlinecite{BH67} that the ratio $B_{5,\vi}^\hnc/B_{5,\co}^\py$ is not a constant.

In conclusion, we expect that the simple relationship \eqref{1.1} can help to gain new insight into the general thermodynamic inconsistency problem, as well as into the connection between the HNC and PY theories. {}From that point of view, the result \eqref{1.1} has a clear pedagogical value, especially taking into account the scarcity of exact and general results in statistical mechanics. On the other hand, from a more practical viewpoint, Eq.\ \eqref{1.1} can be useful to test the correctness of analytical evaluations and/or the accuracy of numerical computations of $B_4$ in the HNC and PY frameworks, especially in the case of mixtures.

\acknowledgments
We are grateful to D. Henderson and A. Giacometti for insightful comments, and to an anonymous referee for suggesting the extension to the multicomponent case. The research of A.S. was supported by the Spanish government through grant No.\ FIS2007-60977, partially financed by FEDER funds, and by the Junta de Extremadura (Spain) through Grant No.\ GRU09038. G.M. is grateful to the Ministerio de Educaci\'on (Spain) for an undergraduate fellowship (beca-colaboraci\'on) during the academic year 2008-2009.

\appendix*

\section{Proof of Eqs.\ \protect\eqref{25}--\protect\eqref{27}}

In this Appendix we derive the identities \eqref{25}--\eqref{27}. For simplicity, we will consider here zero-root diagrams, which are just $V$ times the corresponding one-root diagrams [cf.\ Eq.\ \eqref{2.9}]. Also, we will introduce diagrams such that an arrow on a solid bond from point $i$ to point $j$ represents a term $f_{ij}\rr_{ij}$ while an arrow on a dashed bond represents a term $\nabla_{ij}f_{ij}$. For instance,
\beq
\ytwoBdhd=\int\dd1\int\dd2\int\dd3\int\dd4 f_{13}f_{34}f_{24}f_{14}\rr_{14}\cdot \nabla_{12} f_{12}.
\label{A1}
\eeq

\subsection{Proof of Eq.\ \protect\eqref{25}}
Let us start from
\beq
\ytwoAud=\int\dd1\int\dd2\int\dd3\int\dd4 f_{13}f_{34}f_{24}\rr_{12}\cdot \nabla_{12} f_{12}.
\label{A2}
\eeq
Now, taking into account that $\nabla_{12}f_{12}=-\nabla_2 f_{12}$, we can integrate by parts with the result
\beq
\ytwoAud=-d\ytwoAus+\ytwoAdhv,
\label{A3}
\eeq
where use has been made of the properties $\nabla_2\cdot \rr_{12}=-d$ and $\nabla_2 f_{24}=\nabla_{24}f_{24}$. Moreover, upon writing the second term on the right-hand side of Eq.\ \eqref{A3}, we have exploited the invariance of the zero-root diagrams under rotation and reflection.
Next, making $\rr_{24}=-\rr_{12}+\rr_{13}+\rr_{34}$, one has
\beq
\ytwoAdhv=-\ytwoAud-\ytwoAdhv+\ytwoAdhh.
\label{A4}
\eeq
Integration by parts of the last term on the right-hand side yields
\beq
\ytwoAdhh=-\ytwoAdhv.
\label{A5}
\eeq
{}From Eqs.\ \eqref{A4} and \eqref{A5} one gets
\beq
\ytwoAdhv=-\frac{1}{3}\ytwoAud.
\label{A6}
\eeq
Substitution of Eq.\ \eqref{A6} into Eq.\ \eqref{A3} gives Eq.\ \eqref{25}.

\subsection{Proof of Eq.\ \protect\eqref{26}}
Steps similar to those of Eq.\ \eqref{A3} lead to
\beq
\ytwoBud=-d\ytwoBus+\ytwoBdhv,
\label{A7}
\eeq
\beq
\ytwoCud=-d\ytwoBus-2\ytwoBdhd.
\label{A8}
\eeq
Next, setting $\rr_{14}=\rr_{12}+\rr_{24}$, one has
\beq
\ytwoBdhd=\ytwoBud+\ytwoBdhv.
\label{A9}
\eeq
Combining Eqs.\ \eqref{A7}--\eqref{A9} we get
\beq
\ytwoCud=d\ytwoBus-4\ytwoBdhv.
\label{A10}
\eeq
Finally, Eqs.\ \eqref{A7} and \eqref{A10} gives Eq.\ \eqref{26}.

\subsection{Proof of Eq.\ \protect\eqref{27}}
Again, integration by parts yields
\beq
\ytwoDud=-d\ytwoDus+2\ytwoDdhv.
\label{A11}
\eeq
Also, the equality $\rr_{24}=-\rr_{12}+\rr_{14}$ gives
\beq
\ytwoDdhv=-\ytwoDud-\ytwoDdhv.
\label{A12}
\eeq
Equation \eqref{27} directly follows from Eqs.\ \eqref{A11} and \eqref{A12}.

\end{document}